\documentclass[prd, onecolumn,floats,floatfix,nofootinbib,11pt] {revtex4}
\usepackage{graphicx}
\usepackage{dcolumn}
\usepackage{bm}
\usepackage{graphics}
\usepackage{slashed}
\usepackage{amssymb}
\usepackage{natbib}
\usepackage{amsmath}
\usepackage{url}
\usepackage[usenames,dvipsnames,svgnames,table]{xcolor}
\newcommand{\bea}{\begin{eqnarray}}
\newcommand{\ena}{\end{eqnarray}}
\newcommand{\bean}{\begin{eqnarray*}}
\newcommand{\enan}{\end{eqnarray*}}

\begin{document}

\title{Inspecting Baby Skyrmions with Effective Metrics}
\author{
G. W. Gibbons\\
D.A.M.T.P.,\\  University of Cambridge, U.K.\\
and \\
Laboratoire de Math\'ematiques et de Physique \\
Th\'eorique, Universit\'e de Tours, France 
\\
\texttt{G.W.Gibbons@damtp.cam.ac.uk},\\
E. Goulart\\ 
CAPES/Science without Borders Fellow, Brazil\\
and\\
D.A.M.T.P.,\\  University of Cambridge, U.K.\\\texttt{egoulart@cbpf.br}}
 
\begin{abstract}
In the present paper we investigate the causal structure of the baby Skyrme model using  appropriate geometrical tools. We discuss several features of excitations propagating on top of background solutions and show that the evolution of high frequency waves is governed by a curved effective geometry.  Examples are given for which the effective metric describes the interaction between waves and solitonic solutions such as kinks, antikinks and Hedgehogs. In particular, it is shown how violent processes involving the collisions of solitons and antisolitons may induce metrics which are not globaly hyperbolic.  We argue that it might be illuminating to calculate the effective metric as a diagnostic test for pathological regimes in numerical simulations.
\end{abstract} 

\maketitle

\section{Introduction}

The Skyrme model \cite{skyrme} is a nonlinear theory where the fields take values in a Riemannian  manifold, typically a Lie group. Originaly conceived in the context of high energy physics the model provided a framework where baryons emerged as topologically protected solitons, called skyrmions. Although the relevant field quantities consisted of just $\pi$-meson clouds the fermionic nucleon was obtained as a specific finite-energy, particle-like and stable configuration of the bosonic fields (see \cite{manton} for details). Later, it was realized that the model appears in the low energy effective field theory of quantum chromodynamics in the limit in which the number-of-colours is large \cite{witten}. Recently, apart from nuclear physics, the model proved to be useful in various condensed matter systems, string theory and holographic QCD \cite{multi}, \cite{holo}.

Generically, the associated Euler-Lagrange equations are quasi-linear \cite{tan}-\cite{LAX} and subtle issues arise: i)  wave velocities are not given a priori, but change as functions of initial data, directions of propagation and wave polarization; ii) it is rather usual that the resulting system does not admit solutions which can be continuously extended from the smooth initial configuration: often they tend after a finite time to discontinuous or singular states;  iii) not all initial data are mathematically admissible since for a large class of them the Cauchy problem is ill-posed; iv) there exist the possibility of domains with non-hyperbolic regimes despite the fact that the energy momentum-tensor satisfies the dominant energy condition.   

Systematic treatment on these topics have proven to be difficult and efforts to apply finite difference methods in simulations discovered numerical instabilities. Although general results have been obtained for semilinear wave maps, they do not generalize directly to the quasi-linear regime \cite{shatah}-\cite{tao}. Quite recently, a local well-posedness result was provided by Wong \cite{wong}, sharpening some previous conclusions of Crutchfield \cite{crutch} and Gibbons \cite{gib}.  Unfortunaly, most of the results concerning the time evolution of Skyrme fields depend on numerics or approximation schemes and some of them lack satisfactory analytical explanation.

To gain some intuition into the dynamics of the theory it is convenient to obtain an effective metric description of the linearized waves. This is because the effective metrics uniquely determine the causal structure providing at the same time the natural language to treat the well-posedness of the initial-boundary value problem. As a consequence, they may help in the identification of pathological solutions when they exist and why instabillities appear in some situations involving collisions/scattering of solitons. Among other developments, the effective metric led to the construction of analog models of gravity, which imitate the kinematical properties of gravitational fields, and to insights into the evolutionary properties of hyperbolic PDE's (see \cite{Viss} for a review).

The point of this note is to investigate the causal structure of the baby Skyrme model using the appropriate geometrical tools. The model is a (2+1)-dimensional analogue of the original model with the unit sphere $\mathbb{S}^{2}$ as a target \cite{piette}-\cite{piette2}. Due to the lesser number of dimensions the baby version serves as a toy model for the full theory where some ideas and methods can be tested. They also have physical significance on their own, having applications in some condensed matter systems such as the fractional quantum Hall effect \cite{hall}. In particular, we'll derive a fouth order Fresnel-like equation for the high frequency waves and show that it factorizes into two quadratic characteristic polynomials. It follows that the causal structure of the theory is governed by a duplicity of effective geometries which depend explicitly on the pulled-back geometry for the map. As a consequence, background solutions of the equations behave as a birefringent medium for the linearized waves. We next show that the model admits signature transitions as well, possibly yielding an elliptic regime for the PDE's. In the last section we evaluate the effective metric for some well known solutions such as kinks and baby-skyrmions and show that a wave interacting with the solitons have drastic modifications on its propagation properties.

\section{Baby Skyrme Models: general remarks}

The model is based on a smooth map $\phi^A:M\rightarrow N$ from spacetime $M$ to a target manifold $N$. Here, $(M, \eta)$ is equal to $\mathbb{R}^{1+2}$ equipped with a Minkowski metric $diag(+--)$ and $(N,h)$ is the two-dimensional unit sphere $\mathbb{S}^{2}$ endowed with a positive definite metric $h_{AB}(\phi)$. The lagrangian is constructed from the $(1,1)$ pulled-back tensor $\textbf{L}:= h_{AB}(\phi)\partial^{a}\phi^{A}\partial_{b}\phi^{B}$ and the action is provided by \cite{piette}-\cite{piette2}
\begin{equation}\label{1}
S=\int \frac{1}{2}Tr(\textbf{L})+\frac{\kappa^2}{4}\left[Tr((\textbf{L})^{2})-(Tr(\textbf{L}))^{2}\right]-V(\phi)\ d^{3}x
\end{equation}
where $\kappa^2$ is a parameter controlling the strength of the nonlinearity. The first term in the lagrangian reproduces the classical O(3) sigma model while the second is the three-dimensional analogue of the Skyrme term. While in $(3+1)$ the potential $V(\phi)$ is optional, its presence in the $(2+1)$ model is necessary for the stability of the solitonic solutions\footnote{See \cite{super} for supersymmetric extensions of the model supporting BPS solutions}. 

In what follows we use the conventions: i) capital latin indices, A, stand for the target space quantities and lower latin indices,
a, stand for space-time tensors; ii) $(a,b)=ab+ba$ and $[ab]=ab-ba$ for symetrization and antisymetrization, repectively; iii) $h_{ABPQ}(\phi)\equiv h_{A[P}h_{BQ]}=h_{[AP}h_{B]Q}$. Variation with respect to $\phi^{A}$ yields a system of second order quasi-linear PDE's which can be written in the compact form
\begin{equation}\label{1}
\left(H_{AB}\partial^{a}\phi^{B}\right)_{||C}\partial_{a}\phi^{C}+V_{,A}=0,
\end{equation}
where 
\begin{equation}
H_{AB}(\phi,\partial\phi)\equiv h_{AB}-\kappa^{2} h_{APBQ}\partial^{a}\phi^{P}\partial_{a}\phi^{Q}
\end{equation}
and $||$ represents the covariant derivative with respect to $h_{AB}$, i.e. $h_{AB||C}=0$. In terms of the target connection $\Gamma^{A}_{\phantom a BC}$, (\ref{1}) can be written as
\begin{equation}\label{bse}
\partial_{a}\left(H_{AB}\partial^{a}\phi^{B}\right)-\Gamma^{D}_{\phantom a AC}H_{DB}\partial^{a}\phi^{B}\partial_{a}\phi^{C}+V_{,A}=0,
\end{equation}
which reveals that the equation of motion consists of various types of self-interactions arrising from the non-standard kinetic terms, the potential and the target geometry. Nevertheless, quasi-linearity implies that the system is linear with respect to higher order derivatives of the dependent field variables. Generically, it is possible to express the equations as
\begin{equation}\label{L}
M^{ab}_{\phantom a\phantom a  AB}(\phi,\partial\phi)\ \partial_{a}\partial_{b}\phi^{B}+J_{A}(\phi,\partial\phi)=0,
\end{equation}
where $J_{A}$ stands for semilinear terms in $\phi^{A}$ (lower order derivatives) and the principal symbol is given by
\begin{equation}\label{M}
M^{ab}_{\phantom a\phantom a AB}=\left\{\eta^{ab}h_{AB}+\frac{\kappa^2}{2} h_{APBQ}\left(\partial^{a}\phi^{(Q}\partial^{b}\phi^{P)}-2\eta^{ab}\partial^{c}\phi^{P}\partial_{c}\phi^{Q}\right) \right\}.
\end{equation}

As it is well known, the highest-order terms in derivatives almost completely controls the qualitative behavior of solutions of a partial differential equation. We note that, in this case, the principal symbol $M$ is symmetric with respect to $ab$ and $AB$ i.e. $M^{[ab]}_{\phantom a\phantom a  AB}=M^{ab}_{\phantom a\phantom a  [AB]}=0$. Also, in the limit of $\kappa^2\rightarrow 0$, (\ref{1}) reduces to a semilinear equation
\begin{equation}
\square\phi^{A}+\Gamma^{A}_{\phantom a BC}\partial_{a}\phi^{B}\partial^{a}\phi^{C}=0,
\end{equation}
which repreduces the well-known classical $O(3)$ sigma-model in flat spacetime. 

\section{Causal Structure}

It has been known for some time that the propagation of the excitations of nonlinear field theories in a given background is governed by an effective metric that depends on the background field configuration and on the details of the non-linear dynamics (see \cite{Viss} for a  review). These propagation features can be analysed by means of the eikonal approximation i.e. the regime of small-amplitude/high-frequency waves propagating on top of a smooth solution $\phi^{A}_{0}(x)$. Formally, we consider a one-parameter family of solutions of the form
\begin{equation}
\phi^{A}(x)=\phi^{A}_{0}(x)+\alpha\varphi^{A}(x)exp\left(i\Sigma(x)/\alpha\right),
\end{equation}
and let the real parameter $\alpha\rightarrow 0$. In this limit, only the higher order derivative terms contribute to the propagation laws for the waves. In other words, we can discard source terms $J_{A}$ in (\ref{L}) and consider only the principal part term contributions. 

Defining the wave covector $k_{a}\equiv \partial_{a}\Sigma$, the equation of motion reduces to the eigenvalue equation
\begin{equation}\label{eigenvalue}
\left[M^{ab}_{\phantom a\phantom a AB}(\phi_{0})k_{a}k_{b}\right]\varphi^{B}=0.
\end{equation}
For a general $k_{a}\in T_{x}^{*}M$ we define the symmetric matrix $M_{AB}(\phi_{0},k)\equiv M^{ab}_{\phantom a\phantom a AB}(\phi_{0})k_{a}k_{b}$. It follows that (\ref{eigenvalue}) can be solved only if $k_{a}$ satisfy the algebraic conditions
\begin{equation}
F_{x}(\phi_{0},k)\equiv det(M_{AB}(\phi_{0},k))=0.
\end{equation}
As a consequence, at a given spacetime point, the wave normals are characterized by the roots of a multivariate polynomial of fourth order in $k_{a}$ in the cotangent space. The resulting algebraic variety changes from point to point in a way completely prescribed by the background solution $\phi^{A}_{0}$ and the nonlinearities of the baby Skyrme theory. In general, it will consist at most of two nested sheets, each with the topology of a convex cone. 

The general form of $F_{x}$ is given by a homogeneous polynomial of the form
\begin{equation}\label{fresnel}
F_{x}=det(h_{AB})G^{abcd}(\phi_{0})k_{a}k_{b}k_{c}k_{d}
\end{equation}
with $G^{abcd}$ a completely symmetric quantity. The latter can be written solely in terms of the spacetime metric and the pulled-back geometry. Thus, the vanishing sets of (\ref{fresnel}) constitue the baby-skyrmionic analogues of the Fresnel equation encountered in optics (see \cite{Volker} and references therein for a similar derivation in the context of electrodynamics). They play the role of a fourth order space-time dispersion relation (at least up to a conformal factor).  

Now, the algebraic structure of $G^{abcd}$ reveals that the quartic equation factorizes, yielding the generic birefringence effect, i.e. the characteristic polynomial reduces to a product of two simpler quadratic terms satisfying
\begin{equation}\label{z}
[\eta^{ab}k_{a}k_{b}][(h^{-1})^{cd}k_{c}k_{d}]=0,
\end{equation}
with $h^{-1}$ a reciprocal quadratic form in the cotangent space given by
\begin{equation}\label{E}
(h^{-1})^{ab}\equiv(1-2\kappa^2\mathcal{L}_{s})\eta^{ab}+\left[\kappa^2(1-\kappa^2 Tr(\textbf{L}))L^{ab}+\kappa^{4}L^{ac}L_{c}^{\phantom a b}\right],
\end{equation}
with $\mathcal{L}_{s}$ denoting the lagrangian of the model without the potential term. 

Thus, the wave fronts are not arbitrarily given but satisfy some relations completely prescribed by the pull-back.  In general, $\Sigma(x)$ will solve one quadratic polynomial or the other, although it is possible that there exist some directions where the vanishing sets coincide. Consequently, the model admits two different types of waves. One wave travels with the velocity of light while the other travels with a velocity wich depends implicitly on the solution.

If the quantity $(h^{-1})^{ab}$ is non-degenerate it is possible to define its inverse $h_{ab}$ such that $(h^{-1})^{ac}h_{cb}=\delta^{a}_{\phantom a b}$. The explicit form of $h_{ab}$ may be easily calculated applying the Cayley-Hamilton theorem to the matrix $\textbf{L}$. Because $det(\textbf{L})=0$, it results the matrix relation
\begin{equation}
\textbf{L}^{3}=Tr(\textbf{L})\textbf{L}^{2}+\frac{1}{2}\left[Tr((\textbf{L})^{2})-(Tr(\textbf{L}))^{2}\right]\textbf{L},
\end{equation}
and we obtain
\begin{equation}\label{F}
h_{ab}=(1-2\kappa^2\mathcal{L}_{S})^{-1}\left[\eta_{ab}-\kappa^2 L_{ab}\right].
\end{equation}

In general, the \textit{effective metric} $h_{ab}$ characterizes a Lorentzian metric on spacetime, the null cones of which are the effective ``sound cones" of the theory. The ray vectors associated to the wave fronts are defined as
\begin{eqnarray*}
q^{a}\equiv\eta^{ab}k_{b} \quad &\mbox{if}&\quad \eta^{ab}k_{a}k_{b}=0\\
q^{a}\equiv (h^{-1})^{ab}k_{b} \quad &\mbox{if}& \quad (h^{-1})^{ab}k_{a}k_{b}=0
\end{eqnarray*}
It follows from (\ref{z}) that $q^{a}$ are the vanishing sets of the dual polynomial $G_{x}$, i.e.
\begin{equation}\label{w}
G_{x}(\phi_{0},q)\equiv [\eta_{ab}q^{a}q^{b}][h_{ab}q^{a}q^{b}]=0
\end{equation}
As is well known, these cones completely characterize the causal structure of the theory once a solution is given. Note, however, that (\ref{E}) and (\ref{F}) are defined only up to a conformal transformation $\tilde{h}_{ab}\rightarrow \Omega^{2}(x)h_{ab}$. For the sake of simplicity we assume that $\Omega^{2}=(1-2\kappa^2\mathcal{L}_{S})$ and adopt the reescaled effective structure
\begin{equation}\label{tilde}
\tilde{h}_{ab}=\eta_{ab}-\kappa^2 L_{ab}\quad\quad\quad (\tilde{h}^{-1})^{ab}=\Omega^{-2}h^{ab}.
\end{equation}

The result that the high-energy perturbations of some nonlinear theories propagate along geodesics that are not null in the background geometry but in an effective spacetime has been obtained several times in the literature \cite{ngeo}. This is also true for the baby Skyrme model. Indeed, defining the effective covariant derivative ; such that $\tilde{h}_{ab;c}=0$ it follows that if $\tilde{h}_{ab}q^{a}q^{b}=0$, then
\begin{equation} 
q^{a}_{\phantom a ;b}q^{b}=0,
\end{equation}
which is the equation of a null geodesic. For an arbitrary smooth solution $\phi_{0}$ the quadratic form $\tilde{h}$ is generally curved. Note, however, that it becomes flat in the limit $\kappa^{2}\rightarrow 0$.  Note also that, although different choices $V(\phi)$ may lead to qualitatively different theories, its particular form does not appear explictly in the expression of $\tilde{h}_{ab}$.

\subsection{Signature transitions and breakdowns}
 
 As $\tilde{h}_{ab}$ is a field dependent quantity we cannot guarantee that all roots of $G_{x}$ are automatically real. This will be the case only if the background fields satisfy certain conditions. A direct calculation yields the condition
\begin{equation}
 det(\tilde{h})_{ab}/det(\eta_{ab})=(1-2\kappa^2\mathcal{L}_{S})>0,
\end{equation}
which means that, in order to guarantee that the effective metric has the correct signature $(+--)$, the lagrangian has to satisfy an algebraic constraint. Note that the constraint is the same for all possible potentials. 

Following Manton \cite{MAN} we suppose that $L_{ab}$ can be diagonalized relative to $\eta_{ab}$. It is clear that its eigenvalues are necessarily non-negative so we write them as $L_{ab}=\mbox{diag}(\lambda^{2}_{0},\lambda^{2}_{1},\lambda^{2}_{2})$. A direct calculation yields the components
\begin{equation}\label{decomp}
\tilde{h}_{00}=(1-\kappa^2\lambda_{0}^{2}),\quad\quad\quad \tilde{h}_{11}=-(1+\kappa^2\lambda_{1}^{2}),\quad\quad\quad \tilde{h}_{22}=-(1+\kappa^2\lambda_{2}^{2}).
\end{equation}
It follows that the $00$ component of the effective metric is not always positive, as it should. For sufficiently small values of $\lambda_{0}$, $\tilde{h}$ defines a metric with hyperbolic signature. In this regime, all roots of $\tilde{F}$ and $\tilde{G}$ are real and we have a well defined causal structure. Nevertheless, for large values of $\lambda_{0}$ it is possible to envisage a situation where the effective metric changes its signature. The new regime is characterized by a spacetime region where the equations are of the elliptic type (+,+,+): \textit{instabilities arise}. The two distinct regions are generally separeted by a two-dimensional membrane where the metric is singular. 

One concludes that the physical validity of the baby Skyrme model is constrained by the inequality $\lambda_{0}^{2}<\kappa^{-2}$. For static fields, this is always the case since $\lambda_{0}=0$. For time-dependent fields it is important to keep in mind that a signature transition may occur. If this is case, the system of quasi-linear PDE's looses its physical predictability. This danger is particularly important during violent processes involving the colision of solitons and antisolitons \cite{ANTI}. Therefore, it might be illuminating to calculate the effective metric during numerics as a diagnostic tool. 
 
\section{Examples}

We shall illustrate the formalism described above with examples coming from well known solutions to the baby Skyrme model. In order to explore these aspects, it is convenient to parametrize the fields with spherical coordinates $\phi^{A}=(f,\psi)$ in the target and metric $h_{AB}=\mbox{diag}(1,sin^{2}f)$. From now on we'll work with the so called \textit{old model} potential \cite{broken} which explicitly violates the O(3)-rotational iso-invariance of the theory\footnote{Other possibilites include the \textit{holomorphic} potential and the so called \textit{new model} potential \cite{new}}
\begin{equation}
V(f)=\mu^{2}(1-cosf),
\end{equation}
where $\mu^{2}$ is a mass parameter which we set equal to 1. This potential is analogous to the pion mass term in the $(3 + 1)$ dimensional Skyrme model and is generally associated to external fields in the context of magnetic systems.

\subsection{Sine-Gordon-like solutions}

As shown in \cite{nontop}, any solution of the Sine-Gordon model is also a solution of the baby Skyrme
model. Indeed, in the case of a map with constant azymuthal field ($\partial_{a}\psi=0$) the equation of motion (\ref{bse}) becomes
\begin{equation}\label{sg}
\partial_{a}\partial^{a}f+sinf=0
\end{equation}
which, of course, is the Sine-Gordon equation. Thus, all known solutions of (\ref{sg}) such as kinks, anti-kinks and breathers automatically satisfy the baby Skyrme model equations. Note that, for these types of maps the parameter $\kappa^{2}$ does not appear in the equations. Despite of this fact, the effective metric controlling the causal structure is given by
\begin{equation}
\tilde{h}_{ab}=\eta_{ab}-\kappa^2\partial_{a}f\partial_{b}f.
\end{equation}
We analyzed the causal structure associated to a static kink, a collision of kinks and a collision of kink and anti-kink. The results are illustrated below
\subsubsection{Static Kink}
The kink represents a localized solitary wave, travelling at a velocity $|c|<1$. It is characterized by a solution of the form
\begin{equation}
f(x,t)=4\mbox{arctan}\left(\mbox{exp}\left(\pm\frac{x-x_{0}-ct}{\sqrt{1-c^2}}\right)\right),
\end{equation}
where the $\pm$ signs correspond to localized structures which are called kink and antikink,
respectively. In a reference frame comoving with the solution, we can set $x_{0}=c=0$. A direct calculation yields a diagonal effective geometry of the form
\begin{equation}\label{effg}
\tilde{h}_{00}=1,\quad\quad\tilde{h}_{11}=-(1+4sech^{2}(x)),\quad\quad\tilde{h}_{22}=-1.
\end{equation}
Note that Eqs. (\ref{effg}) are the same for the kink or the antikink solutions, which means that high frequency excitations interact with both solutions in the same way. Obviously, this metric has a Lorentzian signature for all spacetime points.

Fig. 1 shows the lightcones evaluated with the effective metric in the $t-x$ plane. It is seen that the cones concide with those of the Minkowskian geometry far from the kink (antikink), and get thinner inside it, signaling that high-energy perturbations propagate sub-luminally there. We also see that the propagation in the direction $y$ is not affected by the kink (antikink), in accordance with) Eqs. (\ref{effg}). The time spent by a given perturbation inside the localized solution depends on the parameters $\kappa$ and $\mu$, but the solution cannot trap the high-energy perturbations. 
\begin{figure}[1]
       \centering  
       \includegraphics[scale=.4]{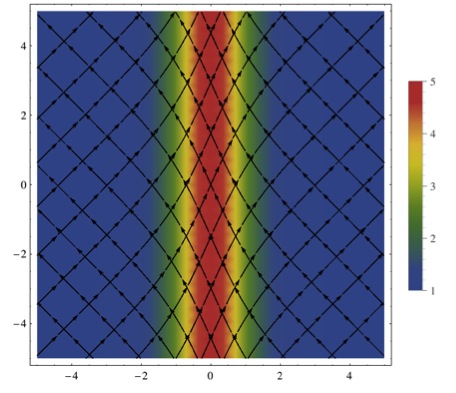}
       \caption{Causal structure of the kink (antikink) solution in the $t-x$ plane. The colors represent the behavior of the function $\sqrt{-\tilde{h}}$ with parameters $\kappa^{2}=1$ and $\mu=1$. In this case it is well defined for all possible spacetime events. Note, however, that it only varies significantly near the kink (antikink), yielding the usual cone for distant excitations.}
       \label{dynamic}
\end{figure}
\begin{figure}[2]
       \centering  
       \includegraphics[scale=0.7]{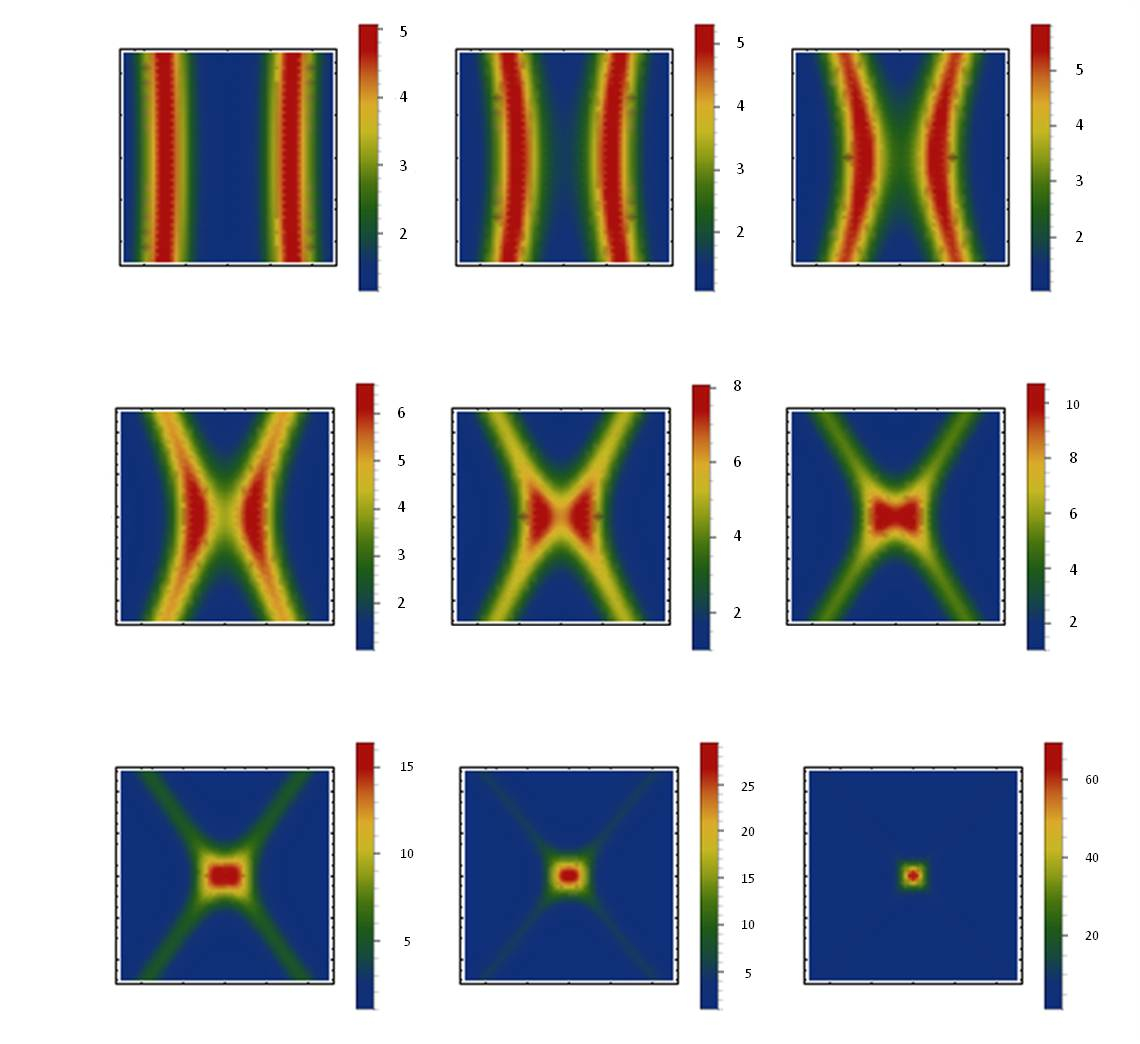}
       \caption{Color diagram representing the function $\sqrt{-\tilde{h}}$ for the kink-kink collision in the $t-x$ plane for $0.1\leq c\leq 0.9$. Note that the function is real for all values of the parameter.}
       \label{dynamic}
\end{figure}
\begin{figure}[H]
       \centering  
       \includegraphics[scale=.6]{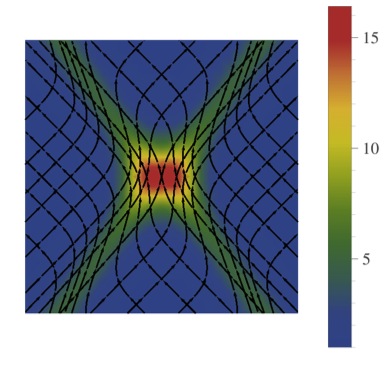}
       \caption{Causal structure associated to the kink-kink collision in the $t-x$ plane for $c=0.7$. Note that the velocities of propagation are drastically modified near the collision.}
       \label{dynamic}
\end{figure} 
\begin{figure}[4]
       \centering  
       \includegraphics[scale=0.6]{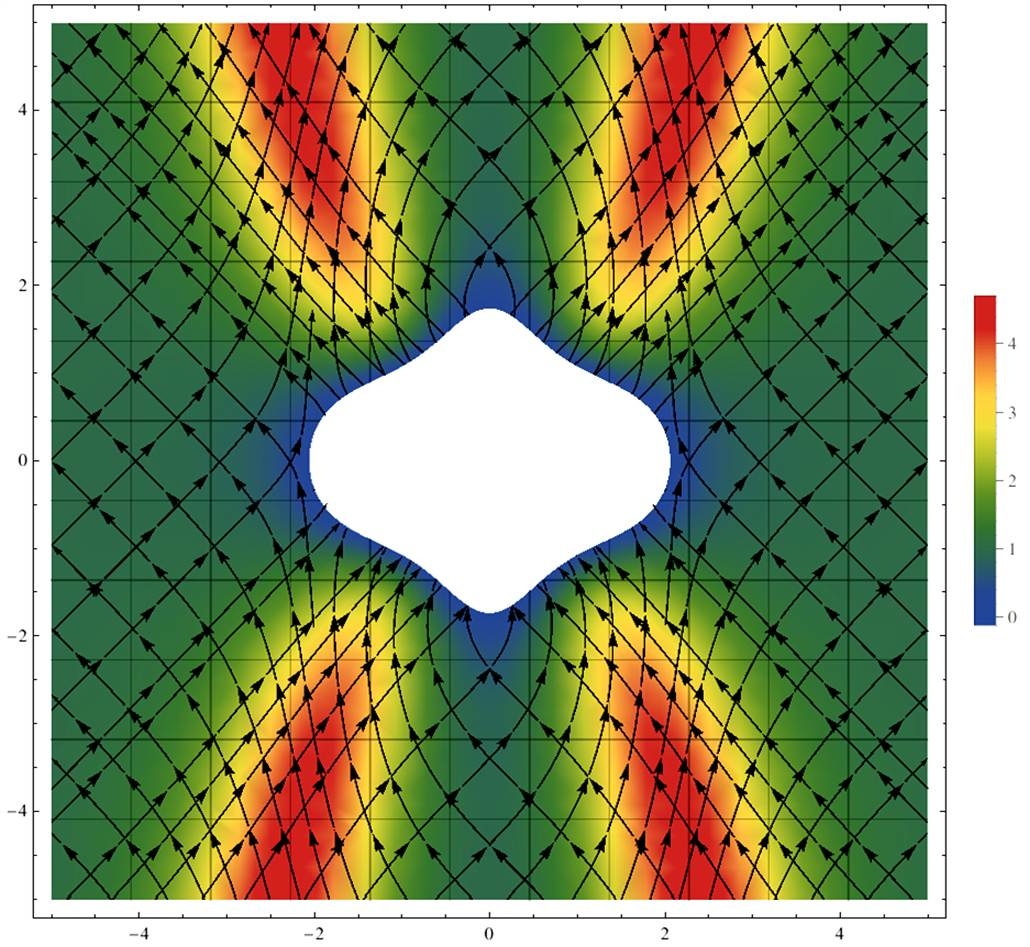}
       \caption{Causal structure of the kink-antikink collision in the $t-x$ plane for $ c= 0.5$. The colors represent the function $\sqrt{-\tilde{h}}$. Note that there exist a region where it vanishes. The white hole in the middle of the figure represents a region with elliptic signature.}
       \label{dynamic}
\end{figure}

\subsubsection{Kink-Kink Collision}

Although there are no static solutions describing multi-solitons in the Sine-Gordon scheme, there are time dependent solutions, which describe the scattering/collisions of two or more kinks \cite{manton}. It is perhaps in the collisions of solitons that the effective geometry may help us in understanding better when instabilities arise and when the solutions are nonphysical. Rather unusually, some of these solutions can be written in analytical form. The kink-kink solution is given by
\begin{equation} 
f(x,t)=4\mbox{arctan}\left(\frac{c\ \mbox{sinh}(\frac{x}{\sqrt{1-c^2}})}{\mbox{cosh}(\frac{ct}{\sqrt{1-c^2}})}\right),
\end{equation}
and describes the collision between two kinks with respective velocities $c$ and $-c$ and approaching the origin from $t\rightarrow -\infty$ and moving away from it with velocities $\pm c$ for $t\rightarrow +\infty$. 

It is well known that there is a repulsive force between two kinks, so, they scatter backwards after the collision. This behavior is reflected in the causal structure of this solution. In Fig. 2 it is ilustrated the behavior of $\sqrt{-\tilde{h}}$ in the $x-t$ plane for different values of the velocity parameter $c$. Again, it is possible to show that the signature of the effective metric is Lorentzian for all possible events, implying that the high frequency limit is well defined. Far away from the kinks, the metric becomes nearly flat and $\tilde{h}\approx 1$. Nevertheless, near the collision, this quantity is considerably modified leading to nontrivial interactions between the waves and the solution. Fig. 3  shows the behavior of the cones for a particular value of $c$.

\subsubsection{Kink-Antikink Collision}
The kink-antikink collision solution is
\begin{equation} 
f(x,t)=4\mbox{arctan}\left(\frac{\mbox{sinh}(\frac{ct}{\sqrt{1-c^2}})}{c\ \mbox{cosh}(\frac{x}{\sqrt{1-c^2}})}\right)
\end{equation}
 with $\pm c$ again describing the velicities of the solitons. As is well known, unlike most topological solitons 
which are annihilated by antisolitons into radiation, 
the kink and antikink scatter elastically. 

From the physical point of view it seems that there is nothing wrong with the kink-antikink collision. Nevertheless, a direct inspection in the effective metric reveals a hidden pathology in the solution. Indeed, the effective metric is not Lorentzian everywhere and becomes singular in a large bidimensional surface in spacetime. This membrane separates the hyperbolic region from a region where the baby Skyrme equations become elliptic (see Fig. 4). Thus, the PDE's are actually mixed in this regime. There is a good chance that the instabilities reported in numerical simulations involving collisions are associated to this signature transition.

\subsection{Hedgehog solutions}

It is well-known that the baby-skyrme model has soliton-like topologically stable static solutions (called baby-skyrmions) and that these solitons can form bound states. To construct these solutions, one must use a radially symmetric ansatz (hedgehog configuration) and reduce the quasi-linear PDE (\ref{bse}) to ordinary differential equations. Adopting polar coordinates $(r,\theta)$  in the spacetime $x-y$ plane,  we consider the class of static solutions of the form
\begin{equation}\label{profile}
 f_{n}=f_{n}(r),\quad\quad\quad \psi=n\theta
\end{equation}
where $f_{n}(r)$ is the so-called profile function and $n\in Z$. We assume also the supplementary conditions $ f_{n}(0)=m\pi$, $(m\in Z)$ and $\lim_{r \to +\infty}f_{n}=0$ to guarantee that the solutions are localized in space and that the total energy is finite.  In analogy with the $(3+1)$ case, the domain of this model is compactified to $\mathbb{S}^2$, yielding the topology required for the classification of its field configurations into homotopy classes $\pi_{2}(\mathbb{S}^{2})=\mathbb{Z}$.
The above ansatz implies that the profile functions $f_{n}$ satisfy the nonlinear ODE\\
\begin{equation}\label{ode}
\left(r+\frac{n^{2}sin^{2}f}{r}\right)f^{''}+\left(1-\frac{n^{2}sin^{2}f}{r^{2}}+\frac{n^{2}f^{'}sinfcosf}{r}\right)f^{'}-\frac{n^{2}sinfcosf}{r}-rsin f=0.
\end{equation}\\
with $f'=df/dr$. Unfortunately, there are no analytic solutions for this model and the baby skyrmion solutions have to be computed numerically. Nevertheless, we may readilly compute the form of the effective geometry. It follows that the causal structure associated to the baby-skyrmions is given by null intervals $\tilde{h}_{ab}dx^{a}dx^{b}=0$ satisfying:\\
\begin{equation}
 dt^2-(1+f_{n}^{'2})dr^2-(r^{2}+n^{2}sin^{2}f_{n})d\theta^2=0,
\end{equation}\\
Note that the effective geometry carries with it the topological degree of the map $n$. In Fig. 5 it is shown the velocity of propagation $v^{2}=dr^{2}/dt^{2}$ as a funtion of the coordinate $r$ for $n=1$ and $n=2$. It would be an interesting task to calculate the effective metrics in simulations of collisions of baby-skyrmions.

\begin{figure}[H]
       \centering  
       \includegraphics[scale=.6]{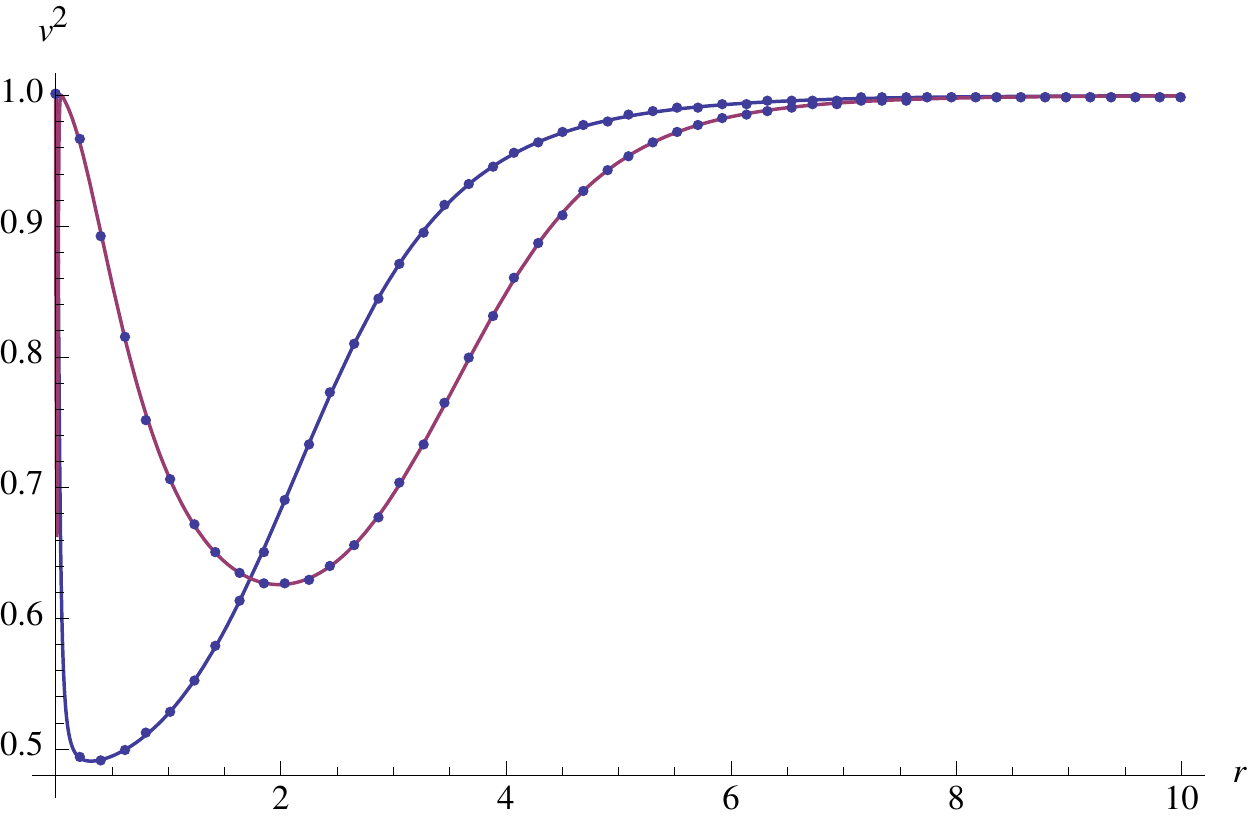}
       \caption{Radial velocity in terms of $r$ for the first two baby-skyrmions $n=1$ (blue curve) and $n=2$ (red curve). Note that for $r\rightarrow \infty$ and $r=0$ the velocities coincide with the velocity of light.}
       \label{dynamic}
\end{figure}

\end{document}